\documentclass{article}
\usepackage[utf8]{inputenc}
\usepackage{cite}
\usepackage{authblk}
\usepackage{setspace}
\usepackage[margin=1in]{geometry}
\usepackage{subcaption}
\usepackage{amsmath,amssymb,amsfonts}
\usepackage{algorithmic}
\usepackage{graphicx}
\usepackage{textcomp}

\graphicspath{{./figures/}}

\title{Region of Interest focused MRI to Synthetic CT Translation using Regression and Classification Multi-task Network}

\author[1,8*]{Sandeep S Kaushik}
\author[2]{Mikael Bylund}
\author[1]{Cristina Cozzini}
\author[3]{Dattesh Shanbhag}
\author[4]{Steven F Petit}
\author[5]{Jonathan J Wyatt}
\author[1,6]{Marion I Menzel}
\author[1]{Carolin Pirkl}
\author[3]{Bhairav Mehta}
\author[7]{Vikas Chauhan}
\author[7]{Kesavadas Chandrasekharan}
\author[2]{Joakim Jonsson}
\author[2]{Tufve Nyholm}
\author[1]{Florian Wiesinger}
\author[8]{Bjoern Menze}

\affil[1]{GE Healthcare, Munich, Germany.}
\affil[2]{Department of Radiation Sciences, Umeå University, Umea, Sweden.}
\affil[3]{GE Healthcare, Bangalore, India.}
\affil[4]{Department of Radiotherapy, Erasmus MC Cancer Institute, Rotterdam, The Netherlands.}
\affil[5]{Translational and Clinical Research Institute, Newcastle University and Northern Centre for Cancer Care, Newcastle upon Tyne Hospitals NHS Foundation Trust.}
\affil[6]{Dept. of Physics, Technical University of Munich, Munich, Germany.}
\affil[7]{Sree Chitra Tirunal Institute of Medical Sciences and Technology (SCTIMST), Trivandrum, India}
\affil[8]{Department of Quantitative Biomedicine, University of Zurich, Zurich, Switzerland}
\affil[*]{Corresponding author. Email: sandeep.kaushik@ge.com}

\date{}

\begin{document}

\maketitle

\begin{abstract}
Synthesizing accurate CT like images is an important step in MR-only clinical workflow. In this work, we present a method for synthetic CT (sCT) generation from zero-echo-time (ZTE) MRI aimed at structural and quantitative accuracies of the image, with a particular focus on the accurate bone density value prediction. We propose a loss function that favors a spatially sparse region in the image. We harness the ability of a multi-task network to produce correlated outputs as a framework to enable localization of region of interest (RoI) via classification, emphasize regression of values within RoI and still retain the overall accuracy via global regression. The network is optimized by a composite loss function that combines a dedicated loss from each task. We demonstrate that the proposed method, despite its architectural simplicity, offers an advantage over other configurations of the network and other popular image generation methods to achieve higher accuracy of performance at a modest computational requirements. This is relevant to sCT where failure to accurately estimate high Hounsfield Unit values of bone could lead to impaired accuracy in clinical applications. We compare the dose calculation maps from the proposed sCT and the real CT in a radiation therapy treatment planning setup.
\end{abstract}

\vspace{2pc}
\noindent{\it Keywords}: MRI Radiation Therapy, Synthetic CT, Multi-task Network, image translation, PET/MR

\section{Introduction}
The electron density information of tissues in the body is essential for accurate dose calculation in radiation therapy treatment planning (RTP) and to compute attenuation correction maps in PET imaging (PET-AC). In the current clinical practice for radiation therapy (RT) treatment planning and in PET/CT imaging, an auxiliary CT image is acquired to provide the necessary electron density information. MRI, with its superior soft-tissue information content, is the preferred imaging modality for tumor and organs-at-risk delineation in RTP \cite{Dirix_2014, Mayr_1993}. This makes it necessary to acquire both CT and MR images in a typical RTP workflow. Image acquisition from two different modalities leads to an extended scan time and incurs an additional image processing step in the clinical workflow to align the information from the two images. In the recent years, there is a growing interest in adapting to an MR-only clinical workflow to leverage the benefits of enhanced soft-tissue contrast in MR images \cite{TNyholm_2019,Tyagi_2017,Chandarana_2018}. To replace a CT image in the RT dose calculation and in PET attenuation correction, an equivalent tissue electron density map needs to be inferred from MRI \cite{Edmund_2017,Moller_2009,Eilertsen_2008}. It is computed by synthesizing a CT like image containing Hounsfield Unit (HU) values corresponding to an MR image, called a synthetic CT (sCT) image. The sCT image generation also finds application in the fields of (a) PET/MR attenuation correction which requires an electron density map generated from an MR image for PET attenuation correction; (b) in MR bone imaging for transcranial focused ultrasound and musculoskeletal applications \cite{Miller_2015, Wiesinger_2016}.

\subsection{Related work}
Some of the traditional methods for generating a synthetic CT image involved bulk tissue density assignment as discussed in \cite{Eilertsen_2008} and image registration between CT and MRI \cite{Kraus_2017, Hsu_2013}. Many recent works have proposed deep CNN learning methods for continuous value sCT generation using variants of U-Net or GAN based regression \cite{Yang_2020, Baydoun_2021, Touati_2021}. Some of the recent reviews on the topic \cite{WangT_2020, SpadeaMF_MasperoM_2021} have extensively summarized related works. The methods summarized in these surveys use one or multiple clinical (T1- or T2-weighted) MR sequences and generate sCT with comparable appearance and quantitative accuracy. Most methods proposed for continuous value sCT synthesis focus on structurally CT-looking images with focus on the overall image accuracy, or global accuracy within separate tissue classes of the body region \cite{GuptaD_2020}. Unpaired image translation using GANs is an active research area, with advantages that include - not requiring co-registered training data and inherent immunity to minor image differences between MR and CT scans. However, concerns about generative inaccuracies and model tractability exist. This makes paired image translation using pixel-wise losses in a supervised framework suitable for a relatively faster adoption in clinical applications for its tractability\cite{LiY_2020}. A well registered cohort of MRI and CT image pairs enables pursuit of supervised image translation methods.

\paragraph{Motivation} In this work, we focus on a specific aspect of MRI to sCT image translation - the accuracy improvement within the bone region in an sCT image. Incorrect estimation of pixel values in the higher electron density regions can lead to higher errors in dose planning when such a misclassified region is in the path of treatment. A reliable bone value assignment in sCT is thus crucial for a reliable MR-only RT workflow since large inaccuracies in bone localization and value estimation could lead to a range of errors in dose calculation in a RT treatment planning as reviewed in \cite{Johnstone_2017}, especially when a tumor is located close to bone or when bone is in the path of radiation (skull) as in case of brain tumors. The bone values in skull range from about $250$ to about $3000$ HU, but only occupy a fractional volume in a typical head CT- about $14$\% of body and $4$\% of image volume at $250$HU. At $900$HU, where the bone density distribution peak is observed, it drops to about $5$\% of body and about $1$\% of image volume. Moreover, a lower error in the estimation of soft-tissue regions can result in a lower error within the body region, overshadowing the inaccuracies in the bone regions due to the tissue bulk of the tissue regions. This motivates our work to consider bone as a region of interest and to develop a solution focused on accurate bone value estimation. Given the large dynamic range and spatial sparsity of the bone regions, a typical model trained on the reference CT image is biased towards the spatially dominant values from soft tissue and background regions, resulting in a reduced accuracy within bone regions. High density bones which are sparser contribute even less towards network optimization.

\paragraph{Our goal} This objective necessitates an evaluation scheme that reflects the accuracy of the synthesized CT across different electron densities and tissue classes as described in \cite{Edmund_2017,Dinkla_2019}. Our goal is to generate an accurate \textsl{synthetic}CT that corresponds to a real CT image over the entire value range, with particular focus on accuracy of the bone regions. We look at two aspects of the solution: $1$) a learning metric that allows to overcome the bias from the dominant region of the image and focus on a specific region of the image such as a \textit{RoI focused loss} as a driving factor and $2$) a framework that offers flexibility over multiple facets of the output, as with a \textit{multi-task network} that is capable of implementing the spatial localization task jointly with the overall HU prediction.

\paragraph{RoI focused loss} If an objective of the model involves sparse data learning, a loss function needs to effectively represent the sparse data and drive the network. In a class unbalanced data, global loss functions are biased towards prominent classes and overshadow the sparser classes. As proposed by \cite{focal_loss}, the problem of class imbalance can be effectively dealt with by down-weighting the prominent class and up-weighting the sparser class. When the RoI in an image is sparse, and hence has weak contribution to learning, it needs a focused enhancement to gain influence on regression network optimization. 

\paragraph{Multi-task network} It was demonstrated by \cite{Caruana_1997} that a network can be trained to perform multiple tasks simultaneously, and that related tasks improve generalization of the network. These works \cite{GuptaD_2020,Duan_2019,Gao_2019} employ a multi-task network to achieve correlated goals which improve the overall solution performance and provide additional outputs from a common network. As noted by them, an optimal convergence of a multi-task network requires training jointly on a composite loss function which comprises of a weighted combination of loss function from each task.

\paragraph{Our contributions} In the context of RoI focused MRI to sCT image translation, 1) we demonstrate that a single task network driven by a global loss is sub-optimal to achieve accurate results; 2) we propose a multi-task network that enables us to introduce a RoI focused loss to separate the global sCT value regression into a local anatomical classification and local HU regression tasks; 3) we demonstrate the efficacy of the joint prediction based on focused loss optimization in an extensive evaluation by comparing sCT generated by other configuration against the real data and by quantitatively evaluating the accuracy of dose calculation from generated sCT against CT in a brain RT treatment planning workflow.
\section{Method}
We aim to build a network capable of mapping a given MR image (${I}_{MR}$) into its corresponding synthetic CT (sCT) image (${I}_{sCT}$) that matches the reference CT image (${I}_{CT}$) values. A CT image can be seen as a combination of three distinct and disjoint electron density regions. i.e, $I_{CT} = (I_{air} \cup I_{tissue} \cup I_{bone})$. In a CT image, air regions are in the HU range [-1000,-400], soft-tissue regions in HU [-250,250], and bone regions in HU [250,3000]. For a given pair of spatially aligned ${I}_{MR}$ (thus $I_{sCT}$) and ${I}_{CT}$, the error between the reference CT and the estimated sCT values can be defined as $e = I_{CT} - I_{sCT}$. A small misestimation of a pixel value results in a lower value of $e$. This typically reflects as pixel value assigned incorrectly, but not necessarily assigned to a different tissue class. However, a large value of $e$ can result in a pixel assigned with an incorrect value which can place it in a different class, thus a misclassification of the pixel. Such an error is more likely to occur around the anatomical class boundary and necessitates a tissue classification in addition to regression. Thus, the HU mapping error $e$ can be seen to comprise of both - classification (inter-class prediction) error and regression (intra-class prediction) error. The overall objective of the network is to map $I_{MR} \rightarrow I_{sCT}$ by minimizing the error $e$.

\subsection{Multi-task Network with Weighted Focus Losses} By separating the tasks of classification and regression, and by optimizing the network to reduce both errors simultaneously, implicit reinforcement can be achieved towards each of the correlated tasks \cite{Caruana_1997}. Although the tasks are correlated, the network is expected to learn them differently from one another. In order to optimize the tasks individually, each one needs to be driven by a dedicated loss function. 

In this work, we propose a framework where a network is assigned three tasks- (a) whole image regression (b) accurate classification of the RoI, and (c) image value regression within the RoI. Each task is driven by a loss function which is tailored to minimize a specific error, thus contributing towards the overall optimal state of the network.

\paragraph{RoI focused loss} The mean absolute error (MAE) is a widely preferred loss function for image regression. However, it is a global measure which neither accounts for imbalance between regional volumes of each class in the image, nor allows to focus on a region of the image as needed. An intuitive way to introduce RoI focus in it is by up-weighting the loss of a region over the rest of the image. The relative volume of a region can be used as an implicit weight factor. For a given region $k$ with volume $N_k$ voxels, MAE within the region is calculated as 
\begin{equation}
MAE_k = \frac{1}{N_k} \sum\limits_{i=1}^{N_k} |y_i - \hat{y}_i|
\end{equation}
\noindent where $y_i$ is the true value and $\hat{y}_i$ is the estimated value. The regional MAE estimates error within the RoI ignoring the background. The weighted MAE for an image with two disjoint image regions $\{k, k’\}$ can then be defined as  
\begin{equation}
wMAE_k = \frac{N_{k'}}{N}*MAE_k + \frac{N_{k}}{N}*MAE_{k'}
\label{wmae_eq} 
\end{equation}
\noindent where $N_k + N_{k'}=N$ is the volume of the entire image. In a scenario of class imbalance where $N_k << N_{k'}$, the value of $MAE_k$ is emphasized by the volume $N_{k'}$, making it comparable to $MAE_{k'}$. This can be seen as focus on a region within the image which is represented by class $k$. 

\paragraph{Dice Loss} For an anatomical classification task, smoothed Dice coefficient is often the preferred loss function \cite{VNet}. Between a given pair of classification probability maps, the Dice loss is defined as 
\begin{equation}
L_D = 1 - ({2 \sum\limits_{i=1}^N x_i\hat{x}_i})/({\sum\limits_{i=1}^N {x_i}^2 + \sum\limits_{i=1}^N {\hat{x}_i}^2})
\end{equation}

\noindent where $x_i$ and $\hat{x}_i$ are the true and predicted bone probability values in the image.

\paragraph{Learning tasks for the multi-task network} In the problem of MR to sCT image translation, with bone being the RoI, the tasks and their losses are assigned as: 
\textbf{(1) synthetic CT image ($I_{sCT}$)}: The primary task of the network is estimation of the entire CT value (HU) range corresponding to different regions in the body. It is driven by the regression loss weighted towards the body region - $L_{body}^{reg} = wMAE_{body}$.
\textbf{(2) bone classification mask ($X_{bone}$)}: This auxiliary task is to classify the bone region within the image. It is intended to regularize the localization of bone regions by penalizing false classification of other regions as bone. This task is driven by the classification loss - $L_{bone}^{class} = L_D$.
\textbf{(3) bone region image ($I_{bone}$)}: This task of the network is intended to generate an image depicting the continuous HU values within the classified bone region. Although it is a subset of the first task, given the wide dynamic range of bone, this loss is meant to explicitly drive the regression of bone values. To focus on the bone region, the rest of the body regions, along with the background are ignored. It is defined by - $L_{bone}^{reg} = MAE_{bone}$. 

\paragraph{Composite loss for training} The overall objective of the network is defined by the composite task - $I_{MR} \rightarrow \{I_{sCT}; X_{bone}; I_{bone}\}$. The neural network is optimized by minimizing the composite loss function defined as  
\begin{equation}
L = w_1*L_{body}^{reg} + w_2*L_{bone}^{class} + w_3*L_{bone}^{reg}
\label{eqn:big_l}
\end{equation}

\noindent The loss weight coefficients $w_1$, $w_2$, and $w_3$ can be either chosen empirically depending on the importance of the task it is meant to drive, or by modeling the uncertainty of each task as proposed by \cite{Kendall_2018}.  

\subsection{Implementation} 
\begin{figure*}[!ht]
\begin{center}
\includegraphics[height=14cm]{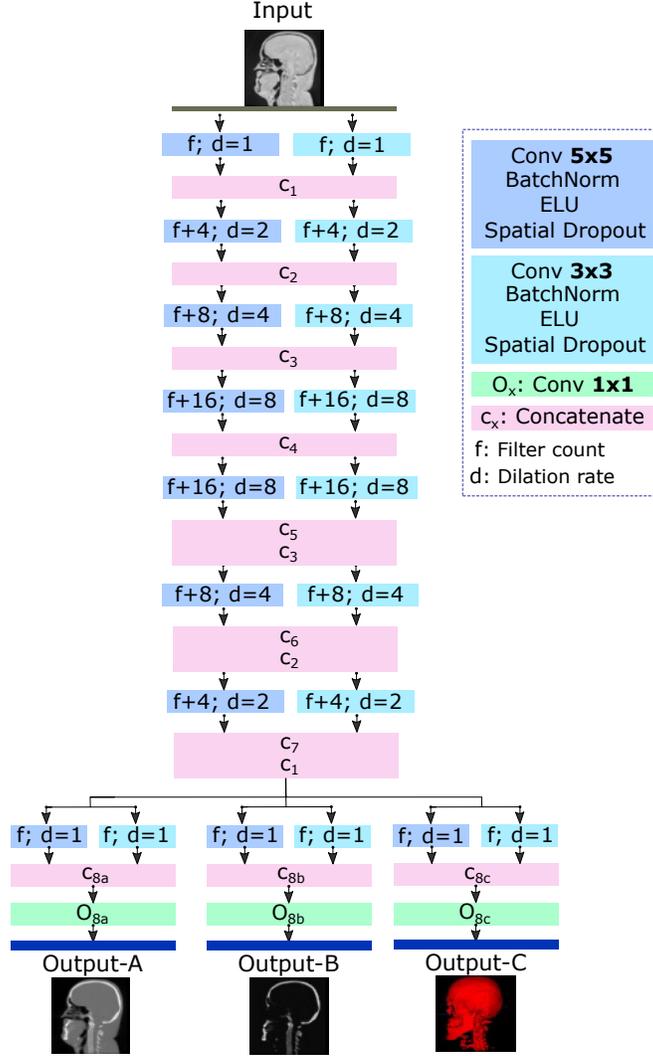}
\end{center}
\caption{Schematic of the DL multi-task network. Each output layer corresponds to a task that the network explicitly learns. Atrous convolution blocks operating at different scales at each level encode high and low level features without compromising the information content due to downsampling.}
\label{fig:dl_arch}
\end{figure*}

\paragraph{Multi-task U-Net architecture} The proposed method was implemented as a $2$D CNN U-Net \cite{UNet} with parallel output layers representing separate tasks. Fig.~\ref{fig:dl_arch} shows the schematic of the DL network architecture. Atrous (dilated) convolutions were used in the network to avoid down-sampling of the image and to preserve sharpness of the translated image. The encoder network consists of $3$-levels x $2$ blocks of $[AtrousConv2D\rightarrow BatchNorm\rightarrow ELU]$ operations. The decoder network consists of $3$-levels x $2$ blocks of $[AtrousConv2D\rightarrow BatchNorm\rightarrow ELU]$. Skip connections were made between corresponding levels of encoder and decoder. The convolution blocks operate at two different scales (filter sizes of $5$x$5$ and $3$x$3$) at each level. This multi-scale feature pyramid is observed to encode the features better than a single block at each level. The decoder path is designed with common shared layers until the final layer. At the final layer, $I_{sCT}$ is obtained via a $[AtrousConv2D\rightarrow BatchNorm\rightarrow Linear]$ to predict HU range [-1000,3000]; $I_{bone}$ is obtained via a $[AtrousConv2D\rightarrow BatchNorm\rightarrow ReLU]$ to predict HU range [250,3000]; and $X_{bone}$ is obtained via a $[AtrousConv2D\rightarrow BatchNorm\rightarrow Sigmoid]$ for a classification mask, each operating with filter size $1$x$1$.  

\paragraph{Loss weights, network outputs, and aggregate sCT} Weights in (\ref{eqn:big_l}) were chosen empirically, by setting the weight of the primary task to unity, and up-weighting the bone segmentation and regression losses. Optimal performance was obtained by empirically setting $w_1=1.0$; $w_2=1.5$; and $w_3=1.3$. The three outputs from the multi-task network are combined to form an aggregate sCT image as the final output. The bone HU image $I_{bone}$ is predicted with the specific objective of quantitative accuracy. Within the bone regions estimated by the mask $X_{bone}$, the HU values of $I_{sCT}$ are replaced with corresponding values from $I_{bone}$. 
\section{Experiments and Results}
\subsection{Dataset and preparation}
The dataset used in the experiments included images acquired from multiple RT treatment planning studies and an MR bone imaging study using a common MR imaging protocol. In the studies from site-A, MR scans were acquired using a GEM HNU surface coil on a 3T, time-of-flight Signa PET/MR scanner (GE Healthcare, Chicago, IL, USA). In the study from site-B, MR scans were acquired using a $32$ channel head coil on a 3T MR scanner (GE Healthcare, Chicago, IL, USA). The proton-density weighted zero echo time (ZTE) MRI presented in \cite{Wiesinger_2016} provides a way to capture robust depiction of bone structures and differentiating bones from air and soft-tissue regions. This MR image provides information required for CT like electron density image synthesis in a single contrast MRI with a fast scan. In total, $54$ brain patients were scanned at $1.5$mm$^3$ isotropic resolution with the ZTE imaging parameters described in \cite{Wiesinger_2018}. For each patient, a CT scan was also acquired which acts as the reference image. All patient studies were approved by respective institutional review boards.
\paragraph{MRI inhomogeneity correction} MR images are prone to B1 field inhomogeneity which results in a multiplicative noise in the form of shading over the image. This was corrected for by applying N4 inhomogeneity correction available in ITK \cite{N4ITK}. The algorithm parameters were chosen empirically to suit the image characteristics based on anatomy, coil, and scan setup.
\paragraph{Image normalization} It is well observed that the performance of neural networks depends on consistency in values across the dataset. The CT image HU values are largely quantitative for a given CT energy source at acquisition. In order to achieve the image value consistency, the ZTE MR images were normalized to their individual z-score value.
\paragraph{Image registration} The real CT image corresponding to each patient MRI serves as the reference image for training the model in a supervised setup. The CT image was aligned to match the MR image space by applying an affine transformation. The registration was performed using the ANTs library \cite{ANTs} by minimizing a combination of mutual-information and cross-correlation metrics. 

\subsection{Experiment setup and evaluation scheme}
Of the total available $54$ cases, $22$ were selected for model training, $12$ for model epoch validation, and $20$ were held out for final model testing. The training sets were augmented by noise introduction, mirroring, rotating, and scaling images at random, resulting in a total of $162$ volumes ($19855$ slices) for training. Validation cases were augmented by noise introduction and mirroring resulting in $36$ volumes ($4860$ slices) for in-training model validation. The network was trained on 2D paired slices of [ZTE, CT]. This method was implemented using functionality available in Tensorflow toolkit on a HP Z$840$ workstation hosting a NVIDIA Titan RTX GPU card. Nadam optimizer with a constant learning rate of 0.001 was used for the model training. A batch size of $15$ was chosen to suit the GPU memory size. As the network is trained to perform three tasks, the convergence was assessed by the minimum combined training loss score from $300$ training epochs. The aggregate sCT volume depicting CT-like HU values is generated from the selected model for a given input ZTE-MR image in the native 3D volume space of the MR image.

The accuracy of the generated sCT within the body region was assessed by comparing against its co-registered CT as the reference. The aspects of evaluation focused on - (a) qualitative accuracy by visual inspection of the sCT image and by pixel-wise difference map; (b) structural accuracy of the bone regions by evaluating the Dice coefficient at multiple HU threshold values, particularly to assess the accurate prediction of high density bones; (c) quantitative HU deviation over the body, bone, tissue, and air regions by measuring MAE within each region; (d) dosimetric calculation for radiation therapy treatment planning.

\subsection{Comparison with other pseudoCT methods}
To evaluate the proposed method in comparison to other works in the literature in context of MRI to sCT image translation, we have chosen two popular methods with widely available implementations - (a) DenseNet \cite{DenseNet} and (a) CycleGAN \cite{CycleGAN}. DenseNet$169$ in a U-Net like framework to provides a baseline of performance. This framework is comparable to the single-task configuration of the proposed method. The CycleGAN framework is setup to work with paired images from our dataset, similar to the methods used in \cite{Li_2020, ShafaiErfani_2019}. Most of the works on MRI to sCT image translation have proposed methods using T1- or T2-weighted MR images which provide better soft tissue accuracy but do not have a signal correlation with bone regions. It results in a better soft-tissue MAE reflecting as a lower MAE in the body region reported those methods despite a higher bone MAE \cite{Dinkla_2019, Yang_2020, LiY_2020}. To the best of our knowledge, there are no publicly available data to benchmark the performance of the methods. The results from these methods are compared with the proposed method visually and quantitatively in the following subsections.

\subsection{Ablation analysis of the learning network}
In order to understand the impact of the proposed focused loss and the multi-task network (3TN), we compared it with the sCT generated from different configurations of the same architecture. (a) To establish a performance baseline of the proposed network architecture, a single task network was trained using the same data as the multi-task network and was driven by a global MAE loss to generate sCT image (1TN$_{GL}$). (b) The single task network configuration of (a) was trained using the weighted MAE loss ~\ref{wmae_eq} for the body region. This network referred to as (1TN$_{FL}$), is intended to study the impact of a weighted loss within a single task. (c) A network configured with two regression tasks - to predict $I_{sCT}$ and $I_{bone}$ (2TN) was trained to understand the impact of absence of the bone classification task.

\paragraph{Experiments with task weights} The effect of varying task weights in the 3TN configuration was studied (i) by setting a constant, relatively higher weighting on the segmentation task and maintaining the other two task weights at unity; (ii) by initializing all the task weights to unity and linearly decreasing the weights on the auxiliary tasks - $I_{sCT}$ and $I_{bone}$. In setting (i), the network at convergence trains to segment the bone region at a marginally higher bone segmentation accuracy with a Dice score of $0.925\pm0.02$ but estimates the pixel values less accurately. That is, the MAE$_{body}$ drops to $80.0\pm10.5$. In setting (ii), the network training drifts from a multi-task configuration performance in the earlier epochs towards a single-task like performance as the training epochs progress.

\subsection{Qualitative analysis} 
\begin{figure*}[!htbp]
\centering
\includegraphics[width=\textwidth]{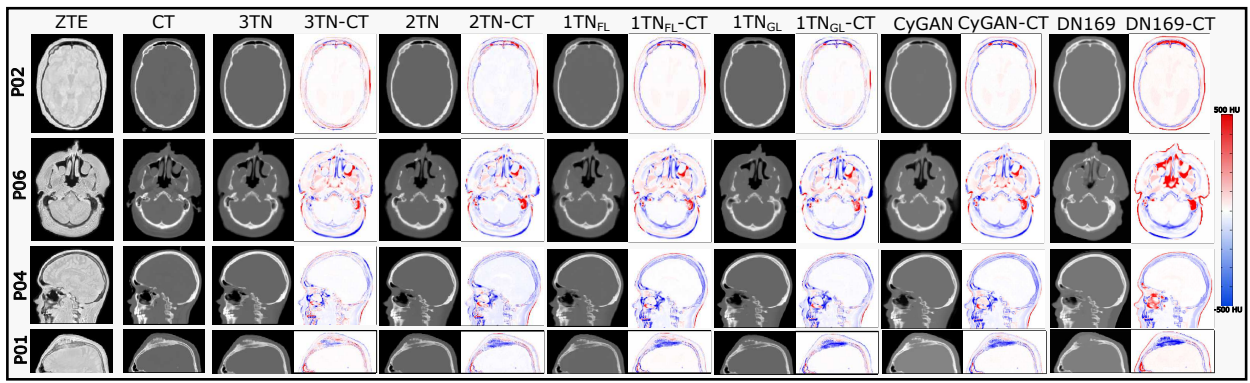}
\caption{Visual assessment of sCT from different schemes in 4 patient cases compared with the corresponding CT. $3TN$: Proposed multi-task network; $2TN$: Two regression tasks network (body and bone regions); $1TN_{FL}$: Single-task network with body focused loss; $1TN_{GL}$: Single-task network with global loss; $CyGAN$: CycleGAN framework in a paired image setting; $DN169$: DenseNet-169 archirecture in a U-Net framework. In the 3TN method, an evident improvement in overall bone value prediction and a better depiction of fine bone structures around the sinus regions can be observed. The depiction of bone pathology in P04 is improved in 3TN compared to other methods. CycleGAN and DenseNet methods focus on the overall image appearance and dominante image information. However their ability to estimate bone and air values are limited and can be seen as higher error in those regions.}
\label{fig:diff_maps}
\end{figure*}

The visual quality of the predicted sCT was assessed by comparing with the reference CT image. Fig.~\ref{fig:diff_maps} shows ZTE-MR image of $4$ different cases from the test dataset along with its co-registered CT. The sCT output from the proposed multi-task network and other configurations from ablation analysis are shown for comparison. The difference map, (sCT$-$CT) is shown alongside each sCT image to highlight the deviations in prediction. The differences seen along the periphery of the body can be attributed to the non-rigid deformation of the body contours between the two scans. Although images from all variants have a similar appearance at first glance and seem to match the CT image, careful observation shows the shortcomings of learning with a global loss function and with one- and two-task networks. The thin bones around the sinus regions in P02 and P06 are either completely missed or their HU values grossly underestimated in other outputs compared to the proposed multi-task network (3TN). A whole image level perspective of P04 further shows that the underestimation of bone values by other configurations are not limited to thin bones around the sinus regions but holds true for skull and teeth regions as well. The image shown in the fourth row of Fig.~\ref{fig:diff_maps} (P01) particularly highlights the effective depiction of bone in presence of a pathology. The intraosseous meningioma of the frontal bone on right side appears as an irregular lobulated hyperdensity in the CT image. This is depicted the closest by the sCT predicted from the proposed 3TN compared to the other configurations as confirmed by the lowest error in the region for sCT from 3TN.

Another way to effectively visualize the impact of RoI-focused learning in prediction of high density bone regions is by comparing the bone depiction in a 3D rendering of the skull as shown in Fig.~\ref{fig:dice_plot}(a). Bone regions from the CT and sCT images were segmented by setting a higher threshold at $600$HU instead of the usual bone threshold at $250$HU. As seen by the skull image from CT, this renders a full skull region including the facial bones, the teeth and the jaw area. Due to the underestimation of bone HU values by 1TN, CycleGAN, and DenseNet169 models, a substantial portion of bones are masked out at the threshold. Another aspect worth noticing in the sCT skull rendering is the structural similarity of depiction of skull abnormality to match the CT image.
\subsection{Quantitative analysis} The validation of quantitative accuracy of the predicted sCT HU values was done by calculating the mean absolute error (MAE) in regions of different electron densities and over the entire body region. Table~\ref{tab1} shows the evaluation performed over the image volume in $20$ test cases within different regions of the image. The MAE in the body region is an indication of overall accuracy of prediction. MAE in the air range HU:[$-1000$,$-400$] indicates the accuracy of prediction within the body air pockets. MAE in the soft-tissue range HU:[$-250$,$250$] indicates the accuracy of prediction in the low density tissue regions. MAE in the range HU:[$250$,$3000$] indicates the accuracy of HU value prediction in the bone regions. The advantage of the proposed method is emphasized particularly by the lowest bone region error. Consequently, better bone accuracy improves air accuracy. This can be attributed to the bone classification task which accurately differentiates air regions from bone as both appear dark in the input MR image. The accuracy of prediction of bone by our proposed method is on par with or better than other U-Net and GAN methods compared in \cite{WangT_2020,SpadeaMF_MasperoM_2021,GuptaD_2020,LiY_2020,Emami_2019}.
\begin{figure}[!b]
\begin{center}
\includegraphics[width=\columnwidth]{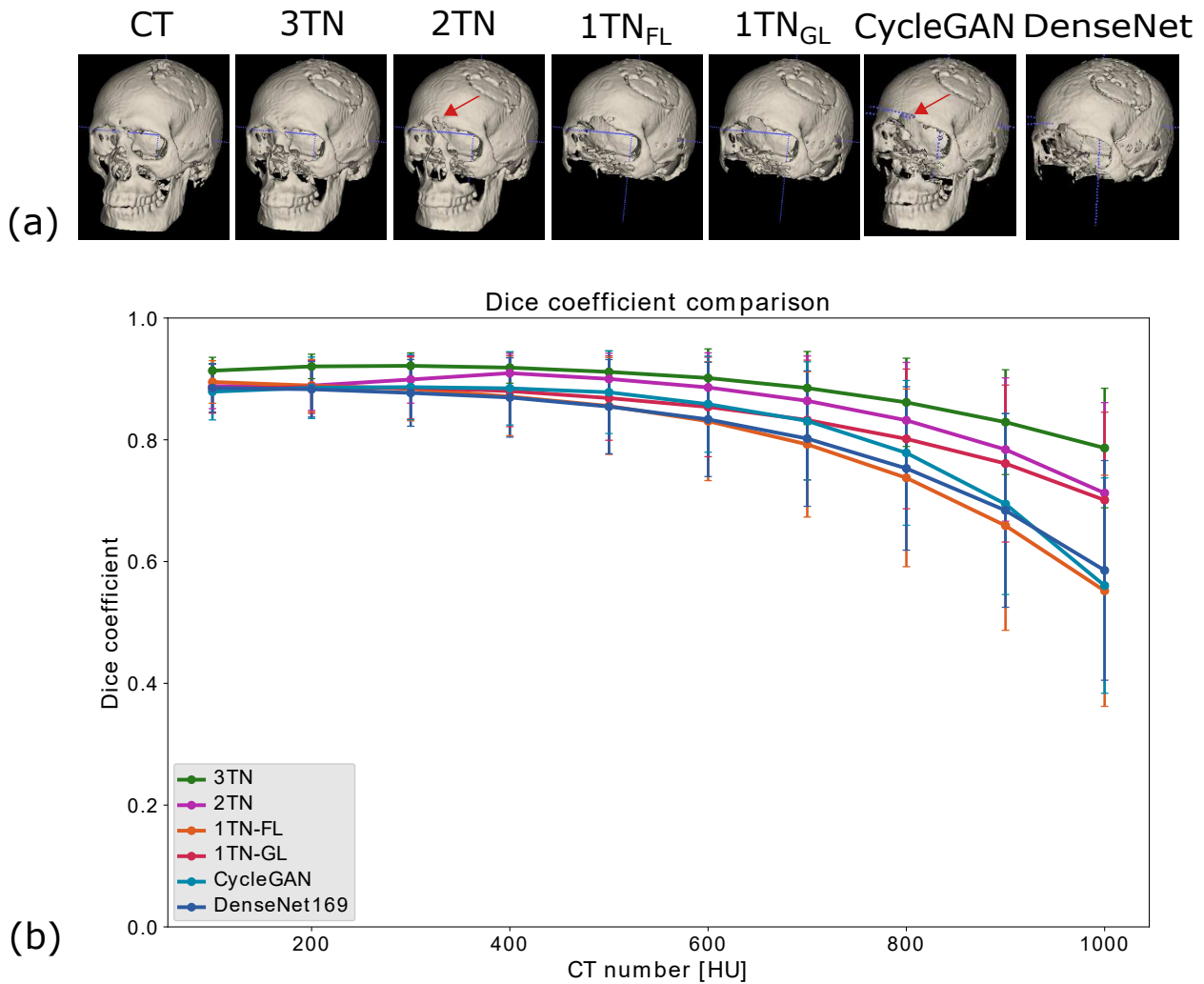}
\end{center}
\caption{(a) Bone depiction in sCT from different configurations compared against CT with threshold set to 600 HU. This shows how the high density bone regions are underestimated by one-task networks with focused and global losses, DenseNet methods, but are well depicted with a focused learning of the bone regions. Although the CycleGAN output estimates the jaw and teeth regions, a large number of missing parts can be observed in the anterior region. (b) Dice coefficient of sCT bone regions at different bone density threshold in different cases shows how all the methods have a higher Dice score at lower HU threshold but decrease at higher threshold. This further confirms the underestimation of high-density bone values.}
\label{fig:dice_plot}
\end{figure}
\begin{table}[b!]
\caption{Region-wise quantitative comparison of 20 sCT image volumes from the proposed method (3TN) and other network configurations and methods with the reference CT. The mean absolute errors in different electron density regions indicate the accuracy of HU value estimation by sCT. The values reported are in Hounsfield Unit (HU)}\label{tab1}
\begin{center}
\resizebox{\textwidth}{!}{\begin{tabular}{|l|c|c|c|c|}
\hline
Method & MAE$_{body}$ & MAE$_{bone}$ & MAE$_{tissue}$ & MAE$_{air}$ \\
\hline
\pmb{Proposed method} &  \pmb{70.0$\pm$8.6} & \pmb{132.0$\pm$14.9} & 36.9$\pm$5.9 & \pmb{174.2$\pm$34.4} \\
Two-task network &  83.7$\pm$11.2 & 166.2$\pm$30.3 & 47.2$\pm$8.4 & 190.5$\pm$34.7 \\
One-task network with body focused loss &  82.9$\pm$10.9 & 183.2$\pm$36.6 & \pmb{34.6$\pm$5.6} & 197.2$\pm$37.0 \\
One-task network with global loss &  84.4$\pm$11.9 & 211.4$\pm$37.9 & 41.6$\pm$7.4 & 193.5$\pm$35.9 \\
CycleGAN &  83.4$\pm$14.8 & 183.8$\pm$38.6 & 35.8$\pm$7.2 & 194.8$\pm$38.6 \\
DenseNet169 &  95.2$\pm$12.9 & 194.9$\pm$40.3 & 38.9$\pm$6.4 & 267.5$\pm$81.6 \\
\hline
\end{tabular}}
\end{center}
\end{table}

The classification accuracy of different sCT regions were assessed by Dice coefficient metric of each region. Table~\ref{tab_dice} confirms that the $3$TN is indeed better in overall classification of regions. The comparable classification from all variants shows the robustness of baseline network performance. However, the bone threshold of $250$ HU does not reflect the accuracy of bone electron density value estimation. Fig.~\ref{fig:dice_plot}(b) shows the Dice coefficient variation at different thresholds of bone densities from $20$ test images. The high Dice score from the proposed method compared to other methods at high bone values can be attributed to the advantage of bone classification and RoI focused loss driving the bone value regression.

\begin{table}[b!]
\caption{Region-wise comparison of Dice coefficient over 20 sCT image volumes from the proposed method (3TN) and other network configurations and methods with the reference CT. This indicates the accuracy of classification of different regions in sCT.}\label{tab_dice}
\begin{center}
\resizebox{\columnwidth}{!}{\begin{tabular}{|l|c|c|c|c|}
\hline
Method & Dice$_{body}$ & Dice$_{bone}$ & Dice$_{tissue}$ & Dice$_{air}$ \\
\hline
\pmb{Proposed method} &  \pmb{0.984$\pm$0} & \pmb{0.918$\pm$0.02} & \pmb{0.957$\pm$0.01} & \pmb{0.809$\pm$0.04} \\
Two-task network &  0.978$\pm$0.01 & 0.892$\pm$0.04 & 0.951$\pm$0.01 & 0.796$\pm$0.04 \\
One-task network with focused loss &  0.974$\pm$0.01 & 0.892$\pm$0.04 & 0.952$\pm$0.01 & 0.80$\pm$0.04 \\
One-task network with global loss &  0.983$\pm$0 & 0.889$\pm$0.04 & 0.951$\pm$0.01 & 0.794$\pm$0.04 \\
CycleGAN &  0.981$\pm$0.01 & 0.875$\pm$0.06 & 0.950$\pm$0.01 & 0.743$\pm$0.07 \\
DenseNet169 &  0.969$\pm$0.01 & 0.864$\pm$0.05 & 0.931$\pm$0.02 & 0.721$\pm$0.09 \\
\hline
\end{tabular}}
\end{center}
\end{table}

\subsection{Dosimetric evaluation}
\begin{figure}[!ht]
\begin{center}
\includegraphics[height=10cm]{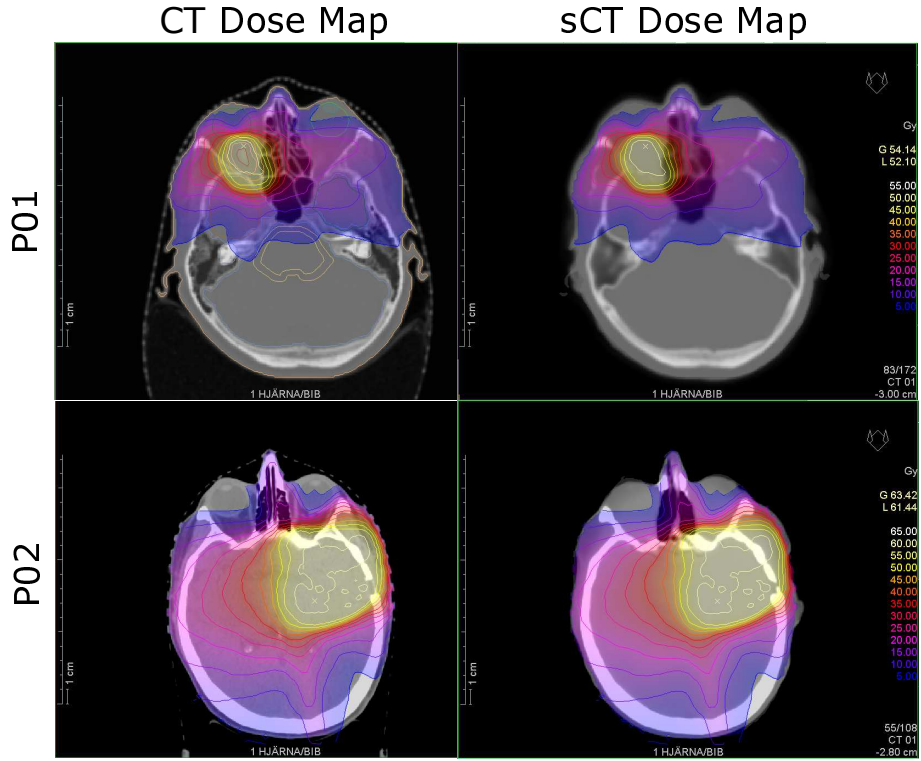}
\end{center}
\caption{Comparison of dose distribution map between CT and sCT based planning. The similar appearance of dose iso-contours indicate a comparable dose reaching the planned target volume (PTV).}
\label{fig:dose_map}
\end{figure}

In order to evaluate the sCT dosimetric performance in RT treatment planning, a comparative analysis was performed with the treatment planning system Oncentra (Oncentra, Elekta, Stockholm, Sweden) using CT and MR data collected for $9$ patients with brain tumors. These $9$ cases were a sub-set of the $20$ test cases in the experiment. Treatment plans were developed based on the CTs using standard clinical guidelines and using RoI drawn by physicians. The treatment plans were then evaluated based on both CT and sCT data, and the results were compared. Fig.~\ref{fig:dose_map} shows the dose distribution maps from CT and sCT over a slice containing the target volume from two patients. A near identical dose distribution can be observed from CT and sCT based planning on these cases. Table~\ref{tab_dose} shows near minimum, near maximum, and average dose difference from CT based plan received by the planning target volume (PTV). The difference in average dose to the PTV relative to the prescribed dose was found to be $0.166\pm0.18$\% which lies within a clinically acceptable range of dose difference of $<0.5$\% and comparable with multiple studies in literature \cite{WangT_2020,SpadeaMF_MasperoM_2021,Kazemifar_2019,Tang_2021}
\begin{table}[htbp]
\caption{Percentage dose differences on target volume D98, DAvg, D2 over 9 patient cases. The near minimum D98 and near maximum doses D2 are doses received by 98\% and 2\% of the PTV volume. DAvg is the average dose to PTV.}\label{tab_dose}
\begin{center}
\resizebox{\columnwidth}{!}{\begin{tabular}{|l|c|c|c|c|c|c|c|c|c|c|}
\hline
Dose & P01 & P02 & P03 & P04 & P05 & P06 & P07 & P08 & P09 & Mean$\pm$Std. \\
\hline
D98 &  0.08 & 0.05 & 0.00 & 0.67 & 0.19 & 0.16 & 0.13 & 0.33 & -1.04 & 0.063$\pm$0.43 \\
\pmb{DAvg} &  0.22 & 0.02 & 0.06 & 0.52 & 0.15 & 0.24 & 0.13 & 0.30 & -0.15 & \pmb{0.166$\pm$0.18} \\
D2 &  0.06 & 0.07 & 0.09 & 0.26 & 0.19 & 0.16 & 0.13 & 0.25 & -0.17 & 0.116$\pm$0.12 \\
\hline
\end{tabular}}
\end{center}
\end{table}
\section{Discussion}
The utility of MR imaging in radiation therapy workflow has been of interest over the past few decades for therapy monitoring \cite{Just_1991} and for tumor depiction and delineation \cite{Mayr_1993}. In recent times, the increased interest in MR-only radiation therapy workflow has led to an active research in the areas of fully automated tumor and organs-at-risk segmentation on MR to aid in therapy planning \cite{Stanescu_2008} \cite{Savenije_2020} \cite{Korte_2021}. The electron density information required in therapy planning which is traditionally provided by a CT image is being replaced with MR derived synthetic CT (sCT) \cite{Kim_2015}. 

In this work, we proposed a novel method for sCT generation from brain MR images with a specific focus on accuracy of bone value estimation. The ZTE MRI provides images with a fast scan time and a single, proton-density weighted contrast which is suitable for bone imaging and sCT generation. The atrous convolution architecture presented here is developed to avoid the image down-sampling in intermediate layers and retain sharpness in the synthesized image. The choice of tasks were based on their correlation and contribution towards the final goal of sCT generation. We have validated the proposed framework by evaluating the impact of the RoI focused loss and the multi-task components separately. The improvement of tissue and bone value estimation seen in Table.~\ref{tab1} in the single task network driven by the focused loss, over global loss demonstrates the impact of learning focused on the body instead of the entire image. Similarly, focusing on bone region in the two-task network results in improvement of bone value estimation. Adding the classification task enhances the focus further on the bone region for both localization and value estimation. 

After experimenting with varying constant task weights and dynamically decreasing auxiliary task weights, the task weights in this work were chosen empirically for optimal performance of sCT output. However, dynamic selection of task weights discussed in \cite{Liu_2019, Kendall_2018} is an interesting future direction for optimizing the task weights for sCT generation. The dynamic behavior of the model performance with varying task weights indicates that the task weight selection is dependent on the application which defines the auxiliary tasks.

Although the HU MAE based validation of the generated sCT images seems to indicate that they are not a perfect match to the reference CT, it is worth noting that their indented use is not for diagnostic purposes but as a workflow aid in radiation therapy planning or PET/MR attenuation correction. The evaluation of the sCT in a retrospective comparison of RT planning shows a clinical outcome that encourages a clinical evaluation over a larger dataset. Differences in image registration is an aspect that needs to be addressed aptly for an accurate comparison of sCT RT planning retrospectively with a CT based planning. Presence of implants in MRI and CT images continue to be a challenge for both of the modalities and results in unpredictable behavior of the network.
\section{Conclusion}
This work was motivated by a specific problem of improving bone value estimation for the intended sCT generation in MR-only radiation therapy workflow. We have presented a novel loss function that is capable of enhancing the focus on sparse regions in an image dynamically. We have proposed a multi-task network by defining the network tasks to translate a ZTE-MR image into a sCT image by classifying the RoI and emphasizing the bone regression accuracy, yet ensuring an overall sCT value accuracy.  The composite loss comprising of dedicated loss from each task drives the network optimization. We have demonstrated, via an extensive evaluation, the superiority of the proposed method over other popular methods and sub-configurations of the proposed method by comparing sCT from each configuration against the CT image. The framework demonstrated here is based on U-net like architecture for intended the simplicity and tractability of the model, and to have a reasonable compute time. A well registered MRI-CT image pairs enables us to take advantage of a supervised learning framework. However, the multi-task framework remains flexible to adapt to other networks of choice and can leverage the inherent qualities of any network. The quantitative evaluation confirms the accuracy of the sCT image, and the comparison of RT treatment planning dose calculation maps from sCT and CT makes case for a larger clinical evaluation of this method.
%
%
%

%

\section*{Acknowledgment}
This research is part of the Deep MR-only Radiation Therapy activity that has received funding from EIT Health. EIT Health is supported by the European Institute of Innovation and Technology (EIT), a body of the European Union receives support from the European Union's Horizon $2020$ Research and innovation program.

%

%
%


\begin{thebibliography}{00}
\bibitem{Dirix_2014}
Dirix P, Haustermans K, Vandecaveye V, The Value of Magnetic Resonance Imaging for Radiotherapy Planning, Seminars in Radiation Oncology, 2014, 24(3):151-159

\bibitem{Mayr_1993}
Mayr N.A, Tali E.T, Yuh W.T, Brown B.P, Wen B.C, Buller R.E, Anderson B, Hussey D.H, Cervical cancer: application of MR imaging in radiation therapy, Radiology, 1993; 189(2):601-608

\bibitem{Miller_2015}
Miller, W. MR bone imaging. Journal of Therapeutic Ultrasound vol. 3,Suppl 1 O37. 30 Jun. 2015, doi:10.1186/2050-5736-3-S1-O37

\bibitem{TNyholm_2019}
Jonsson J, Nyholm T, Söderkvist K. The rationale for MR-only treatment planning for external radiotherapy Clin Transl Radiat Oncol.;18:60–65. (2019)

\bibitem{Tyagi_2017}
Tyagi, N., Fontenla, S., Zelefsky, M. et al. Clinical workflow for MR-only simulation and planning in prostate. Radiat Oncol 12, 119 (2017). 

\bibitem{Chandarana_2018}
Chandarana H, Wang H, Tijssen RHN, Das IJ. Emerging role of MRI in radiation therapy. J Magn Reson Imaging. 2018 Dec;48(6):1468-1478.

\bibitem{Eilertsen_2008}
Karsten Eilertsen, Line Nilsen Tor Arne Vestad, Oliver Geier and Arne Skretting, A simulation of MRI based dose calculations on the basis of radiotherapy planning CT images, Acta Oncologica, (2008) 47:7, 1294-1302

\bibitem{Edmund_2017}
Edmund, J.M., Nyholm, T. A review of substitute CT generation for MRI-only radiation therapy. Radiat Oncol 12, 28 (2017).

\bibitem{Kraus_2017}
Kraus K.M, Jäkel O, Niebuhr N.I, Pfaffenberger A, Generation of synthetic CT data using patient specific daily MR image data and image registration, Phys. Med. Biol. 2017, 62 1358

\bibitem{Hsu_2013}
Hsu S-H, Cao Y, Huang K, Feng M, Balter J.M, Investigation of a method for generating synthetic CT models from MRI scans of the head and neck for radiation therapy, Phys. Med. Biol. 2013, 58 8419

\bibitem{Moller_2009}
Martinez-Möller A1, Souvatzoglou M, Delso G, Bundschuh RA, Chefd'hotel C, Ziegler SI, Navab N, Schwaiger M, Nekolla SG., Tissue classification as a potential approach for attenuation correction in whole-body PET/MRI: evaluation with PET/CT data., J Nucl Med. 2009 Apr;50(4):520-6	

\bibitem{WangT_2020}
Wang T, Lei Y, Fu Y, Wynne JF, Curran WJ, Liu T, Yang X, A review on medical imaging synthesis using deep learning and its clinical applications, J Appl Clin Med Phys 2021; 22:1:11–36

\bibitem{SpadeaMF_MasperoM_2021}
Spadea MF, Maspero M, Zaffino P, Seco J, Deep learning-based synthetic-CT generation in radiotherapy and PET: a review, Medical Physics 2021; Vol.48 Issue.11, Pages 6537-6566

\bibitem{Yang_2020}
Yang H, Sun J, Carass A, Zhao C, Lee J, Prince J.L., Xu Z, Unsupervised MR-to-CT Synthesis Using Structure-Constrained CycleGAN, IEEE Transactions on Medical Imaging, vol. 39, no. 12, pp. 4249-4261, Dec. 2020

\bibitem{Baydoun_2021}
Baydoun A, Xu KE, Heo JU, Yang H, Zhou F, Bethell LA, Fredman ET, Ellis RJ, Podder TK, Traughber MS, Paspulati RM, Qian P, Traughber BJ, Muzic RF. Synthetic CT Generation of the Pelvis in Patients With Cervical Cancer: A Single Input Approach Using Generative Adversarial Network. IEEE Access. 2021;9:17208-17221. 

\bibitem{Touati_2021}
Touati R, Trung Le W, Kadoury S, A feature invariant generative adversarial network for head and neck MRI/CT image synthesis, Phys. Med. Biol., Volume 66, Number 9, 2021 

\bibitem{LiY_2020}
Li Y, Li W, Xiong J, Xia J, Xie Y, Comparison of Supervised and Unsupervised Deep Learning Methods for Medical Image Synthesis between Computed Tomography and Magnetic Resonance Images, BioMed Research International, vol. 2020, Article ID 5193707, 9 pages, 2020.

\bibitem{GuptaD_2020}
Gupta D, Kim M, Vineberg KA, Balter JM, Generation of Synthetic CT Images From MRI for Treatment Planning and Patient Positioning Using a 3-Channel U-Net Trained on Sagittal Images, Front. Oncol., Vol 9., Sep. 2019

\bibitem{Leynes_2018}
Leynes AP, Yang J, Wiesinger F, Kaushik SS, Shanbhag DD, Seo Y, Hope TA, Larson PEZ., Zero-Echo-Time and Dixon Deep Pseudo-CT (ZeDD CT), J Nucl Med. 2018 May;59(5):852-858.

\bibitem{Dinkla_2019}
Dinkla, A.M., Florkow, M.C., Maspero, M., Savenije, M.H.F., Zijlstra, F., Doornaert, P.A.H., van, Stralen, M., Philippens, M.E.P., van den, Berg, C.A.T. and Seevinck, P.R., Dosimetric evaluation of synthetic CT for head and neck radiotherapy generated by a patch‐based three‐dimensional convolutional neural network. Med. Phys.,2019, 46: 4095-4104

\bibitem{Maspero_2018}
Maspero M, Savenije MHF, Dinkla AM, Seevinck PR, Intven MPW, Jurgenliemk-Schulz IM, Kerkmeijer LGW, van den Berg CAT., Dose evaluation of fast synthetic-CT generation using a generative adversarial network for general pelvis MR-only radiotherapy., Phys Med Biol. 2018 Sep 10;63(18):185001

\bibitem{Kazemifar_2019}
Kazemifar S, McGuire S, Timmerman R, Wardak Z, Nguyen D, Park Y, Jiang S, Owrangi A., MRI-only brain radiotherapy: Assessing the dosimetric accuracy of synthetic CT images generated using a deep learning approach., Radiother Oncol. 2019 Jul;136:56-63

\bibitem{Zeng_2019}
Zeng G., Zheng G. (2019) Hybrid Generative Adversarial Networks for Deep MR to CT Synthesis Using Unpaired Data. In: Shen D. et al. (eds) Medical Image Computing and Computer Assisted Intervention – MICCAI 2019. MICCAI 2019. Lecture Notes in Computer Science, vol 11767. Springer, Cham

\bibitem{Johnstone_2017}
Johnstone E, Wyatt JJ, Henry AM, Short SC, Sebag-Montefiore D, Murray L, Kelly CG, McCallum HM, Speight R, A systematic review of synthetic CT generation methodologies for use in MRI-only radiotherapy, International Journal of Radiation Oncology Biology Physics (2017)

\bibitem{Caruana_1997}
Caruana, R. Multitask Learning. Machine Learning 28, 41–75 (1997).

\bibitem{Duan_2019}
Duan J., Bello G., Schlemper J., Bai W., Dawes T.J.W., Biffi C., de Marvao A., Doumou G., O’Regan D.P., Rueckert D., Automatic 3D Bi-Ventricular Segmentation of Cardiac Images by a Shape-Refined Multi- Task Deep Learning Approach, in IEEE Transactions on Medical Imaging, vol. 38, no. 9, pp. 2151-2164, Sept. 2019.

\bibitem{Gao_2019}
Gao F., Yoon H, Wu T, Chu X, A feature transfer enabled multi-task deep learning model on medical imaging, Expert Systems with Applications, Volume 143, 2020

\bibitem{focal_loss}
Lin T, Goyal P., Girshick R., He K., Dollár P., Focal Loss for Dense Object Detection, ICCV 2017, pp. 2999-3007.

\bibitem{VNet}
Milletari F., Navab N., Ahmadi S., V-Net: Fully Convolutional Neural Networks for Volumetric Medical Image Segmentation, 2016 Fourth International Conference on 3D Vision (3DV), 2016, pp. 565-571.

\bibitem{UNet}
Ronneberger O., Fischer P., Brox T. (2015) U-Net: Convolutional Networks for Biomedical Image Segmentation. In: Navab N., Hornegger J., Wells W., Frangi A. (eds) Medical Image Computing and Computer-Assisted Intervention – MICCAI 2015. MICCAI 2015. Lecture Notes in Computer Science, vol 9351. Springer, Cham

\bibitem{Wiesinger_2018}
Wiesinger F, Bylund M, Yang J, Kaushik S, Shanbhag D, Ahn S, Jonsson JH, Lundman JA, Hope T, Nyholm T, Larson P, Cozzini C., Zero TE-based pseudo-CT image conversion in the head and its application in PET/MR attenuation correction and MR-guided radiation therapy planning., Magn Reson Med. 2018 Oct;80(4):1440-1451

\bibitem{Wiesinger_2016}
Wiesinger F., Sacolick L.I., Menini A., Kaushik S.S., Ahn S., Veit-Haibach P., Delso G., and Shanbhag D.D, Zero TE MR Bone Imaging in the Head, Magnetic Resonance in Medicine 75:107–114 (2016)

\bibitem{ANTs}
Avants, B. B., Tustison, N. J., Stauffer, M., Song, G., Wu, B., Gee, J. C., The Insight ToolKit image registration framework, Frontiers in Neuroinformatics, 8, 44. (2014)

\bibitem{N4ITK}
Tustison NJ, Avants BB, Cook PA, Zheng Y, Egan A, Yushkevich PA, Gee JC. N4ITK: improved N3 bias correction. IEEE Trans Med Imaging. 2010 Jun;29(6):1310-20. 

\bibitem{Emami_2019}
Emami, H., Dong, M., Nejad‐Davarani, S.P. and Glide‐Hurst, C.K. (2018), Generating synthetic CTs from magnetic resonance images using generative adversarial networks. Med. Phys., 45: 3627-3636.

\bibitem{Kazemifar_2019}
Kazemifar, S., McGuire, S., Timmerman, R., Wardak, Z., Nguyen, D., Park, Y., Jiang, S., Owrangi, A., MRI-only brain radiotherapy: Assessing the dosimetric accuracy of synthetic CT images generated using a deep learning approach, Radiotherapy and Oncology, Volume 136, 2019, Pages 56-63

\bibitem{Tang_2021}
Tang, B., Wu, F., Fu, Y., Wang, X., Wang, P., Orlandini, L.C., Li, J. and Hou, Q. (2021), Dosimetric evaluation of synthetic CT image generated using a neural network for MR‐only brain radiotherapy. J Appl Clin Med Phys.

\bibitem{DenseNet}
Huang G, Liu Z, v. d. Maaten L, Weinberger K.Q., Densely Connected Convolutional Networks, CVPR 2017, pp. 2261-2269.

\bibitem{CycleGAN}
Zhu, J-Y., Park, T., Isola, P., and Efros, A.A., "Unpaired Image-to-Image Translation using Cycle-Consistent Adversarial Networks", in IEEE International Conference on Computer Vision (ICCV), 2017.

\bibitem{Kendall_2018}
Kendall A, Gal Y, Cipolla R. Multi-Task Learning Using Uncertainty to Weigh Losses for Scene Geometry and Semantics, CVPR 2018, pp. 7482–7491

\bibitem{Just_1991}
Just M, Kahaly G, Higer H.P, Rösler H.P, Kutzner J, Beyer J, Thelen M, Graves ophthalmopathy: role of MR imaging in radiation therapy, Radiology, 1991; 179(1):187-190

\bibitem{Stanescu_2008}
Stanescu T, Jans H-S, Pervez N, Stavrev P, Fallone B.G, A study on the magnetic resonance imaging (MRI)-based radiation treatment planning of intracranial lesions, Phys. Med. Biol. 2008; 53 3579

\bibitem{Savenije_2020}
Savenije MHF, Maspero M, Sikkes GG, van der Voort van Zyp JRN, T J Kotte AN, Bol GH, T van den Berg CA. Clinical implementation of MRI-based organs-at-risk auto-segmentation with convolutional networks for prostate radiotherapy. Radiat Oncol. 2020 May 11;15(1):104. 

\bibitem{Korte_2021}
Korte, JC, Hardcastle, N, Ng, SP, Clark, B, Kron, T, Jackson, P. Cascaded deep learning-based auto-segmentation for head and neck cancer patients: Organs at risk on T2-weighted magnetic resonance imaging. Med. Phys. 2021; 00 1– 16. 

\bibitem{Kim_2015}
Kim J, Glide-Hurst C, Doemer A, Wen N, Movsas B, Chetty I.J, Implementation of a Novel Algorithm For Generating Synthetic CT Images From Magnetic Resonance Imaging Data Sets for Prostate Cancer Radiation Therapy, International Journal of Radiation Oncology*Biology*Physics, Vol. 91, Iss. 1, 2015

\bibitem{Liu_2019}
Liu S, Liang Y, and Gitter A, Loss-Balanced Task Weighting to Reduce Negative Transfer in Multi-Task Learning. Proceedings of the AAAI Conference on Artificial  Intelligence 2019, 33(01), 9977-9978.

\bibitem{Kendall_2018}
Kendall A, Gal Y, and Cipolla R, Multi-task learning using uncertainty to weigh losses for scene geometry and semantics. In Proceedings of the IEEE conference on computer vision and pattern recognition 2018 (pp. 7482-7491).

\bibitem{Li_2020}
Li W, Li Y, Qin W, Liang X, Xu J, Xiong J, et al. Magnetic resonance image (MRI) synthesis from brain computed tomography (CT) images based on deep learning methods for magnetic
resonance (MR)-guided radiotherapy. Quant Imaging Med Surg 2020;10:1223–36. https://doi.org/10.21037/qims-19-885

\bibitem{ShafaiErfani_2019}
Shafai-Erfani G, Lei Y, Liu Y, Wang Y, Wang T, Zhong J, et al. MRI-Based Proton Treatment Planning for Base of Skull Tumors. International Journal of Particle Therapy 2019;6:12–25.
https://doi.org/10.14338/IJPT-19-00062.1.

\bibitem{LiY_2020}
Li Y, Li W, Xiong J, Xia J, Xie Y. Comparison of Supervised and Unsupervised Deep Learning Methods for Medical Image Synthesis between Computed Tomography and Magnetic Resonance Images. Biomed Res Int. 2020;2020:1-9.

\end{thebibliography}
\end{document}